\documentclass[aps,prd,reprint,superscriptaddress,floatfix]{revtex4}
\usepackage{graphicx}
\usepackage{amsmath}
\usepackage{amssymb}
\usepackage{booktabs}
\usepackage{xcolor}
\usepackage{bm}
\usepackage{multirow}
\usepackage{makecell}
\usepackage[colorlinks=true,citecolor=cyan,urlcolor=blue,bookmarks=true,bookmarksopen=true,bookmarksnumbered=true,bookmarksopenlevel=3]{hyperref}

\begin{document}

\title{FIMPs in a two-component dark matter model with $Z_2 \times Z_4$
symmetry}

\author{\textsc{XinXin Qi}}
\email{qxx@dlut.edu.cn}
\affiliation{Institute of Theoretical Physics, School of Physics, Dalian University of Technology, No.2 Linggong Road, Dalian, Liaoning, 116024, P.R.China }
\author{\textsc{Hao Sun}}
\email{haosun@dlut.edu.cn}
\affiliation{Institute of Theoretical Physics, School of Physics, Dalian University of Technology, No.2 Linggong Road, Dalian, Liaoning, 116024, P.R.China }

\begin{abstract}
We investigate the FIMP-FIMP regime in a two-component dark matter model with a $Z_2\times Z_4$ symmetry, where a singlet scalar $S$ and a Majorana fermion $\chi$ serve as the dark matter candidates. A singlet scalar $S_0$ with vacuum expectation value $v_0$ generates the fermion mass through the relation $m_\chi=y_{sf}v_0$. We show that the tiny Yukawa coupling $y_{sf}$ needed to reproduce the observed relic abundance naturally leads to a large symmetry-breaking scale $v_0$, which induces an ultra-feeble portal coupling $\lambda_{ds}$ responsible for the production of $S$.
We find that $\lambda_{ds}$ can reach values of $10^{-25}\lesssim\lambda_{ds}\lesssim10^{-13}$, while gravitational freeze-in provides an irreducible contribution at extremely small couplings. Our results demonstrate that the relic abundance constraint, combined with symmetry breaking and freeze-in dynamics, naturally drives the portal interaction responsible for  scalar dark matter production  into the ultra-feeble regime.

\end{abstract}
\maketitle
\setcounter{footnote}{0}
\section{Introduction}
\label{sec:intro}

The nature of dark matter (DM) and its interaction with the Standard Model (SM) remain among the most important open questions in particle physics and cosmology. Although the weakly interacting massive particle (WIMP) paradigm provides an appealing explanation for the observed relic abundance through thermal freeze-out, the absence of confirmed signals from direct and indirect detection experiments ~\cite{Planck:2018vyg} has motivated the exploration of dark matter scenarios with much weaker interactions. The freeze-in mechanism ~\cite{Bernal:2017kxu} offers an alternative possibility, where dark matter particles are produced through extremely feeble interactions with the thermal bath and never reach thermal equilibrium~\cite{Hall:2009bx}.

In freeze-in scenarios, the tiny couplings required to reproduce the observed relic abundance are often introduced as free parameters. This raises an important question: can ultra-feeble dark matter interactions emerge naturally from the structure of a dark sector rather than being imposed by hand? Multi-component dark matter models~\cite{Boehm:2003ha,Zurek:2008qg,Profumo:2009tb,Pandey:2017quk,Belanger:2020hyh,Belanger:2022esk,Barger:2008jx,Liu:2011aa,Qi:2024zkr,Bhattacharya:2016ysw,Bhattacharya:2017fid,Bhattacharya:2022qck,Sakharov:1994pr,Qi:2025jpm,Khlopov:2021xnw,DiazSaez:2021pmg,DiazSaez:2023wli,Borah:2024emz,Costa:2022oaa,Costa:2022lpy,Khan:2024biq,Choi:2021yps,Sheng:2026aro,Bhattacharya:2024mZN} provide a suitable framework to investigate this possibility, since different dark matter components may have distinct production mechanisms and interaction strengths. The interplay between multiple dark sector particles can therefore generate non-trivial relations among masses, couplings, and cosmological production processes.

In this work, we investigate the FIMP-FIMP regime of the $Z_2\times Z_4$ model introduced in Ref.~\cite{Qi:2025jpm}, where a singlet scalar $S$ and a Majorana fermion $\chi$ are the DM candidates. Previous studies of this framework have explored the WIMP-WIMP and mixed WIMP-FIMP scenarios~\cite{Qi:2025jpm,Qi:2025znq}. Here, we focus on the fully freeze-in regime and investigate the origin of the required ultra-feeble interaction hierarchy. A key feature of the model is that $\chi$ acquires its mass via $m_\chi=y_{sf}v_0$ after the singlet $S_0$ develops a vev. We show that the small Yukawa coupling $y_{sf}$ required for the freeze-in production of $\chi$ naturally leads to a large symmetry-breaking scale $v_0$, which subsequently induces an ultra-feeble portal coupling $\lambda_{ds}$ controlling the production of the scalar DM component $S$.

We perform a systematic analysis of the FIMP-FIMP parameter space and study the relic abundance and phenomenological implications of the resulting ultra-feeble interactions. We find that the portal coupling $\lambda_{ds}$ can reach values of $10^{-25}\lesssim\lambda_{ds}\lesssim10^{-13}$, while gravitational freeze-in provides an irreducible contribution at extremely small couplings. Our results demonstrate that the relic abundance constraint, combined with symmetry breaking and freeze-in dynamics, naturally drives the dark matter interactions into the ultra-feeble regime.

The paper is organized as follows. Section~\ref{sec:2} presents the model. Section~\ref{sec:fimp} discusses the FIMP production mechanism and Boltzmann equations. Section~\ref{sec:discussion} presents the viable parameter space, naturalness analysis, and comparison with the WIMP-WIMP and mixed regimes. Section~\ref{sec:sum} summarizes.

\section{Model description}\label{sec:2}
In this section, we present a two-component dark matter model with $Z_2 \times Z_4$ symmetry. We extend the SM by introducing two singlet scalars $S$ and $S_0$ together with one Majorana fermion $\chi$, where $S$ and $\chi$ are the DM candidates and $S_0$ acquires a non-zero vacuum expectation value  (vev) $v_0$. The charges carried by the particles in the model are listed as follows:
\begin{table}[htbp]
\center
 \begin{tabular}{|l|r|}
 \hline
 Particle  & $Z_2 \times Z_4$ \\
 \hline
 $\mathrm{SM}$    & (1,1)\\
 \hline
 $S$     & (-1,1)\\
 \hline
 $S_0$ & (1,-1)\\
 \hline
 $\chi$ & (1,i)\\
 \hline
  \end{tabular}
  \caption{ The charges of the particles  under $Z_2\times Z_4$ symmetry.}
  \label{table1}
\end{table}
The new Lagrangian is therefore given by:
\begin{align}\label{eq:Lnew}
 \mathcal{L}_{new}  &\supset \frac{1}{2}M_1^2 S^2 + \frac{1}{4}\lambda_{s}S^4-\frac{1}{2} \mu_0^2S_0^2+\frac{1}{4}\lambda_{0} S_0^4 - \mu_H^2|H|^2 +\lambda_{H}|H|^4
  +\lambda_{dh}S^2|H|^2 + \lambda_{ds}S^2S_0^2 \notag\\
 +& \lambda_{sh}S_0^2|H|^2 + y_{sf}S_0\chi^{T}\chi
 \end{align}
 where $H$ is the SM Higgs doublet. The mass parameters $\mu_0^2$, $\mu_H^2$, and $M_1^2$ are all taken to be positive. With our sign conventions, the scalar potential is $\mathcal{V} = -\mathcal{L}_{\rm scalar}$, under which $-\frac{1}{2}\mu_0^2 S_0^2$ and $-\mu_H^2|H|^2$ provide the tachyonic mass terms that drive spontaneous symmetry breaking, while $+\frac{1}{2}M_1^2 S^2$ supplies a positive mass-squared contribution for the DM candidate $S$. In the unitary gauge, $H$ and $S_0$ can be expressed as:
    \begin{equation}
H=\left(\begin{array}{c} 0 \\ \frac{v+h}{\sqrt{2}}\end{array} \right) \, , \quad
S_0=s_0+ v_0\, ,\quad
\end{equation}
 where $v =246$ GeV corresponds to the electroweak symmetry breaking vev and $v_0$ is the vev of $S_0$. After spontaneous symmetry breaking (SSB), the masses of $S$ and $\chi$ are given by:
 \begin{eqnarray}
 m_S^2= M_1^2 +2\lambda_{ds}v_0^2 +\lambda_{dh}v^2, ~~m_{\chi}=y_{sf}v_0,
 \end{eqnarray}
  where $m_S$ ($m_{\chi}$) represents the mass of $S$ ($\chi$). The squared mass matrix mixing $s_0$ and $h$ reads:
  \begin{eqnarray}
    \mathcal{M}= \left(
    \begin{array}{cc}
     2\lambda_{0}v_0^2 & \lambda_{sh}vv_0 \\
     \lambda_{sh} vv_0 & 2\lambda_{H}v^2 \\
    \end{array}
    \right).
  \end{eqnarray}
The physical masses of the two Higgs states $h_1, h_2$ are then given by
\begin{align}
\label{Higgsmass}
	m^2_{1} &= \lambda_H v^2 + \lambda_{0} v_0^2 
	- \sqrt{(\lambda_H v^2 - \lambda_{0} v_0^2)^2 + (\lambda_{sh}vv_0)^2},\notag\\
  m^2_{2} &= \lambda_H v^2 + \lambda_{0} v_0^2 
	+ \sqrt{(\lambda_H v^2 - \lambda_{0} v_0^2)^2 + (\lambda_{sh}vv_0)^2}
\end{align} 
The mass eigenstates $(h_1, h_2)$ and the gauge eigenstates $(h, s_0)$ are related via
\begin{align}
\label{Higgs mixing}
	\begin{pmatrix}
		h_1 \\ h_2
	\end{pmatrix} = 
	\begin{pmatrix}
    	\cos\theta & -\sin\theta \\
		\sin\theta &  \cos\theta
	\end{pmatrix}
	\begin{pmatrix}
		h\\ s_0
	\end{pmatrix}.
\end{align} 
where
 \begin{eqnarray}
    \tan 2\theta= \frac{\lambda_{sh}vv_0}{\lambda_{0}v_0^2 - \lambda_{H}v^2}
\end{eqnarray}
We identify $h_1$ with the observed 125 GeV SM Higgs boson and $h_2$ as the new Higgs boson. Choosing $m_1$ and $m_2$ as input parameters, the couplings $\lambda_H$, $\lambda_0$, and $\lambda_{sh}$ are given by:
\begin{align}
\label{para_quartic}
	\lambda_H &= 
 		\frac{(m_{1}^2 +m_{2}^2) - 
    	\cos 2 \theta (m_{2}^2 - m_{1}^2)}{4 v^2}, \nonumber\\
	\lambda_{0} &= 
 		\frac{(m_{1}^2 +m_{2}^2) + 
    	\cos 2 \theta (m_{2}^2 - m_{1}^2)}{4 v_0^2} , \\
	\lambda_{sh} &= 
 		\frac{\sin 2 \theta (m_{2}^2 - m_{1}^2)}{2 v v_0}  \nonumber
\end{align}
The mixing angle between the SM Higgs and additional scalars is stringently constrained by NLO corrections to the $W$ boson mass~\cite{Lopez-Val:2014jva}, perturbative unitarity requirements~\cite{Robens:2021rkl}, and direct searches at LEP and the LHC~\cite{CMS:2015hra,Strassler:2006ri}.

Since the $Z_4$ symmetry is spontaneously broken by the vacuum expectation value of $S_0$, domain walls may be generated during the phase transition. We assume that the $Z_4$-breaking phase transition occurs before a subsequent inflationary epoch. The exponential expansion during inflation dilutes the domain wall density and avoids the associated cosmological domain wall problem. This assumption does not affect the freeze-in dynamics considered in this work.

\section{FIMP dark matter} \label{sec:fimp}
 We are interested in the regime where both dark matter species are produced via the freeze-in mechanism. In this regime, the number densities of $\chi$ and $S$ are negligible in the early Universe and never reach thermal equilibrium due to their feeble interactions; their eventual freeze-in determines the observed DM relic density today. 
  \subsection{Thermalization of $h_2$ }\label{sec:sintheta}
The new Higgs boson $h_2$ is assumed to be in thermal equilibrium with the SM bath in the early Universe, a necessary condition for it to serve as the mediator of DM production in the FIMP scenario. This places a stringent lower bound on the mixing angle $\sin\theta$. 
The thermalization of $h_2$ is governed by its interaction rate with SM particles, which for the dominant Higgs portal process scales as $\Gamma_{h_2\leftrightarrow\mathrm{SM}} \sim \sin^2\theta \,\Gamma_{\rm SM}(m_2)$, where $\Gamma_{\rm SM}(m_2)$ represents the total decay width that the SM Higgs boson would have if its mass were $m_2$. The condition for $h_2$ to reach thermal equilibrium is:
\begin{equation}\label{th}
 \langle\Gamma_{h_2\leftrightarrow\mathrm{SM}}\rangle \gtrsim H(T\sim m_2).
\end{equation}
with $\langle\Gamma_{h_2\leftrightarrow\mathrm{SM}}\rangle$ being the thermally averaged decay width.
Roughly speaking, one can estimate that:
\begin{eqnarray}
H(m_2) \approx 1.4\times 10^{-12}\left(\frac{m_2}{1\ {\rm TeV}}\right)^2\ {\rm GeV},
\end{eqnarray}
 \begin{eqnarray}
 \langle\Gamma_{h_2\leftrightarrow\mathrm{SM}}\rangle = \Gamma_{h_2\leftrightarrow\mathrm{SM}}\cdot\frac{K_1(m_2/T)}{K_2(m_2/T)}\mid_{m_2 \sim T} \approx 0.37\ \Gamma_{h_2\leftrightarrow\mathrm{SM}}.
 \end{eqnarray}
Therefore, Eq.~\eqref{th} yields:
\begin{eqnarray}
 0.37 \sin^2\theta \cdot\Gamma_{\rm SM}(m_2) \gtrsim H(m_2).
\end{eqnarray}
Hence we obtain:
\begin{eqnarray}
\sin\theta \gtrsim \sqrt{\frac{H(m_2)}{0.37\Gamma_{SM}(m_2)}}.
\end{eqnarray}
For $m_2=1$~TeV, the lower bound on $\sin\theta$ is approximately $9 \times 10^{-8}$. Moreover, the minimum of $\sin\theta$ satisfies:
\begin{eqnarray}
 \sin\theta_{\rm min} \propto \sqrt{\frac{m_2^2}{m_2^3}} = \frac{1}{\sqrt{m_2}},
\end{eqnarray}
 As $m_2$ varies from 1 TeV to 2 TeV, $\sin\theta_{\rm min}$ changes by only a factor of $\sim 1.4$. Taking a conservative value, we have:
\begin{eqnarray}
 \sin\theta \gtrsim 10^{-7},
 \end{eqnarray}
 which remains valid throughout the range $m_2 \in [300\ {\rm GeV},2\ {\rm TeV}]$.  This lower bound on $\sin\theta$ implies that the contribution of SM particles to $\chi$ production cannot be rendered negligible by arbitrarily fine-tuning $\sin\theta$ to smaller values. If $\sin\theta$ is much smaller, $h_2$ will never reach thermal equilibrium and $\chi$ will instead be generated via the so-called ``mediator-dominated freeze-in'' process~\cite{Konar:2025gvh}, and we will discuss this case in the future work.

Having established the lower bound $\sin\theta \gtrsim 10^{-7}$, we now justify fixing $\sin\theta = 10^{-4}$ in the subsequent analysis. (The upper bound from LHC Higgs signal strength measurements, electroweak precision data, and perturbative unitarity~\cite{Lopez-Val:2014jva,Robens:2021rkl} lies at $\sin\theta \lesssim 10^{-1}$ for $m_2\sim 1$~TeV, leaving a wide allowed window.) First, for $\chi$ production with $m_\chi < m_2/2$, the dominant channel $h_2 \to \chi\chi$ has a decay width $\Gamma \propto y_{sf}^2\cos^2\theta \approx y_{sf}^2$, which is essentially independent of $\sin\theta$ for $\theta \ll 1$. Second, for $S$ production with $m_S < m_2/2$, the $h_2 \to SS$ decay width scales as $|\sin\theta \cdot \lambda_{dh}v + 2\lambda_{ds}v_0|^2$. In the parameter regions of interest, $v_0 \sim 10^9$--$10^{15}$~GeV, and over most of the viable parameter space $\lambda_{ds}v_0$ exceeds $\sin\theta \cdot \lambda_{dh}v$ by several orders of magnitude (marginal corner cases where the two terms are comparable are possible but do not affect the qualitative picture). Consequently, the $S$ decay production rate is largely insensitive to $\sin\theta$ as well. For the scattering-dominated regimes ($m_{\chi,S} > m_2/2$), the SM-initiated channels $XX \to \chi\chi, SS$ scale as $\sin^2\theta$, while the dark-sector channels $h_2h_2 \to \chi\chi, SS$ are $\sin\theta$-independent. Varying $\sin\theta$ within the allowed window changes the relative weight of these two contributions but does not alter the qualitative structure of the viable parameter space: the resulting shift in the required couplings can always be absorbed by a modest rescaling of $y_{sf}$ or $\lambda_{ds}$. The chosen value $\sin\theta = 10^{-4}$ is therefore a representative benchmark that lies well within the allowed window, ensures robust $h_2$ thermalization, and does not affect any of our qualitative conclusions.

\subsection{Boltzmann equations}
 The current dark matter relic density measured by the Planck collaboration is $\Omega_{\rm DM}h^2 = 0.1198 \pm 0.0012$~\cite{Planck:2018vyg}. We consider that both $\chi$ and $S$ are produced via the freeze-in mechanism and contribute to the total DM abundance. The Boltzmann equations for the abundances of $S$
  and $\chi$ are given as follows:
  \begin{align}\label{be1} 
\frac{dY_S}{dx} &= \frac{1}{3H}\frac{ds}{dx}[\langle \sigma v \rangle^{XX \to SS}\bar{Y_X}^2+ \langle \sigma v \rangle^{h_2h_2 \to SS}\bar{Y}_{h_2}^2+\theta(m_2-2m_S)\Gamma_{h2S}\bar{Y}_{h_2}+ \theta(m_1-2m_S)\Gamma_{h1S}\bar{Y}_{h_1}] 
\end{align}

\begin{align}\label{be2}
 \frac{dY_{\chi}}{dx}  =&  \frac{1}{3H}\frac{ds}{dx} [\langle \sigma v \rangle^{XX \to \chi\chi}\bar{Y_X}^2+\langle \sigma v \rangle^{h_2h_2 \to\chi\chi}\bar{Y}_{h_2}^2+ \theta(m_2-2m_{\chi})\Gamma_{h_2\chi}\bar{Y}_{h_2}]. \ \ \ \         
\end{align}
where $x=m_S/T$ with $T$ being temperature, $\theta(x)$ is the Heaviside function, $s$ denotes the entropy density.
  $Y_S$ and $Y_{\chi}$ are the abundances of $S$ and $\chi$ defined by $Y_S \equiv n_S/s$ and $Y_{\chi} \equiv n_{\chi}/s$,
where $n_S$ and $n_{\chi}$ are the number densities of $S$ and $\chi$. $\bar{Y}_{h_1}$ and $\bar{Y}_{h_2}$ are the equilibrium abundances of $h_1$ and $h_2$,
\begin{eqnarray}
\bar{Y}_{h_i}=\frac{45x^2m_i^2}{4\pi^4 g_{*S}m_S^2}K_2\!\left(\frac{m_i}{m_S}x\right),\qquad i=1,2,
\end{eqnarray}
where $K_2(x)$ is the modified Bessel function of the second kind and $g_{*S}$ is the effective number of entropy degrees of freedom.
In Eqs.~\eqref{be1}--\eqref{be2}, the notation $\langle\sigma v\rangle^{XX\to SS}\,\bar{Y}_X^2$ (and similarly for $\chi$) is a shorthand for the sum over all kinematically accessible SM initial states,
\begin{equation}
\langle\sigma v\rangle^{XX\to SS}\,\bar{Y}_X^2 \equiv \sum_{a,b\,\in\,\mathrm{SM}} \langle\sigma v\rangle^{ab\to SS}\,\bar{Y}_a^{\rm eq}\,\bar{Y}_b^{\rm eq},
\end{equation}
where $a,b$ run over SM quarks, leptons, and gauge bosons, and $\bar{Y}_a^{\rm eq}$ is the equilibrium yield of species $a$. In the numerical analysis, this sum is evaluated automatically by \textsc{micrOMEGAs}, which includes all relevant SM degrees of freedom. The single-particle expression $\bar{Y}_X$ with a nominal mass $m_X$ shown in some textbooks is not used in our computation; we retain the compact notation $\bar{Y}_X^2\langle\sigma v\rangle^{XX\to\cdots}$ only for brevity.
$H$ is the Hubble expansion rate of the Universe, and $\langle \sigma v \rangle$ is the thermally averaged annihilation cross section~\cite{Gondolo:1990dk}. $\Gamma_{h_1S}$, $\Gamma_{h_2S}$, and $\Gamma_{h_2\chi}$ denote the
thermally averaged decay rates for $h_1\to SS$, $h_2\to SS$, and $h_2\to\chi\chi$, respectively, defined as~\cite{Zhang:2024sox}:
\begin{eqnarray}\label{h2d}
\Gamma_{h_1S}=\Gamma_{h_1 \to SS}\frac{K_1(m_1/T)}{K_2(m_1/T)},\Gamma_{h_2S}=\Gamma_{h_2 \to SS}\frac{K_1(m_2/T)}{K_2(m_2/T)},\Gamma_{h_2\chi}=\Gamma_{h_2 \to \chi\chi}\frac{K_1(m_2/T)}{K_2(m_2/T)}.
\end{eqnarray}
with
\begin{eqnarray*}
\Gamma(h_1\to SS) = \frac{|\cos\theta \cdot \lambda_{dh}v - \sin\theta \cdot 2\lambda_{ds}v_0|^2}{32\pi m_1}\sqrt{1-\frac{4m_S^2}{m_1^2}},
\end{eqnarray*}
\begin{eqnarray*}
\Gamma(h_2\to SS) = \frac{|\sin\theta \cdot \lambda_{dh}v + \cos\theta \cdot 2\lambda_{ds}v_0|^2}{32\pi m_2}\sqrt{1-\frac{4m_S^2}{m_2^2}},
\end{eqnarray*}
\begin{eqnarray*}
\Gamma_{h_2 \to \chi\chi}=\frac{y_{sf}^2\cos^2\theta m_2}{4\pi}(1-\frac{4m_{\chi}^2}{m_2^2})^{3/2},
\end{eqnarray*}
 where $K_1(x)$ is the modified Bessel function of the second kind and $\cos\theta\approx 1$ for $\sin\theta\ll 1$. 
 
 Note that in Eq.~\eqref{be2} we have omitted the possible contribution of $h_1\to\chi\chi$ when $m_\chi<m_1/2$. This channel is suppressed by $\sin^2\theta$ and is always subdominant compared to other production processes. As discussed above, we fix $\sin\theta=10^{-4}$ throughout the following analysis; this value, although small, ensures robust $h_2$ thermalization. On the other hand,
 Eqs.~\eqref{be1}--\eqref{be2} also omit elastic co-scattering processes such as $S + X \leftrightarrow S + X$, $\chi + X \leftrightarrow \chi + X$, $S + h_2 \leftrightarrow S + h_2$, and $\chi + h_2 \leftrightarrow \chi + h_2$. Elastic co-scattering modifies only the momentum distribution, not the number density, and is therefore irrelevant for the relic density computation.  Processes of the type $h_2 + X \to SS + X$ and $h_2 + X \to \chi\chi + X$, where a thermal $h_2$ converts to a DM pair through scattering with an SM particle, are $2\to 3$ processes that are subdominant when the two-body decay $h_2 \to SS,\chi\chi$ is kinematically open: the $2\to 3$ cross section carries an extra $\sin^2\theta$ suppression (from the $h_2$--SM vertex) relative to the decay width, giving $\langle\sigma v\rangle_{h_2 X} n_X^{\rm eq}/\Gamma_{h_2} \sim 4\pi\sin^2\theta \sim 10^{-7}$ at $T\sim m_2$. When the decay channel is closed, the $2\to 2$ channels already included in Eqs.~\eqref{be1}--\eqref{be2} ($XX$, $h_2h_2$) capture the leading production mechanisms, and the $2\to 3$ processes are further Boltzmann-suppressed relative to $XX$ for $T\lesssim m_2$.

 \subsection{Numerical analysis}
 In this work, we take the following six parameters as free inputs:
 \begin{eqnarray}
 m_{\chi},y_{sf},m_2,m_S,\lambda_{ds},\lambda_{dh}.
 \end{eqnarray}
Note that in the limit $y_{sf} \to 0$, the model reduces to the two singlet scalar DM case with $S$ as FIMP, while in the limit $\lambda_{ds}\to 0$ and $\lambda_{dh}\to 0$, it reduces to the singlet fermion DM case with $\chi$ as FIMP.  As we will discuss below, $y_{sf}$ is stringently constrained and cannot be arbitrarily small while still satisfying the DM relic density constraint.

In the region $m_{\chi}<m_2/2$, the $\chi$ relic density is generated by the decay $h_2\to\chi\chi$ together with $XX\to\chi\chi$ scattering. One can estimate the upper bound on $y_{sf}$ in this region by switching off the SM contribution, so that $\chi$ is produced entirely by $h_2$ decay. In the limit $m_\chi\ll m_2/2$, the decay width in Eq.~\eqref{h2d} simplifies to:
\begin{eqnarray}
 \Gamma_{h_2 \to \chi\chi} \approx \frac{m_2y_{sf}^2}{4\pi}, \label{eq:gammah2chichi}
\end{eqnarray}
which grows linearly with $m_2$, as expected for a two-body decay mediated by a Yukawa coupling. 
 
 The $\chi$ yield $Y_\chi$ can be computed by solving the Boltzmann equation~\eqref{be2}, which simplifies to:
 \begin{eqnarray}
 sT\frac{dY_{\chi}}{dT}= -\frac{\gamma_{h_2 \to \chi\chi}(T)}{H(T)},
 \end{eqnarray}
 where $H(T)$ is the Hubble expansion rate at temperature $T$ and $\gamma_{h_2\to\chi\chi}(T)$ is the thermally averaged FIMP production rate:
 \begin{eqnarray}
 \gamma_{h_2 \to \chi\chi} =\frac{m_2^2T}{2\pi^2}K_1(m_2/T)\Gamma_{h_2\to \chi\chi},
 \end{eqnarray}
 
 For high temperatures, $T>m_2$, we obtain \cite{Yaguna:2023kyu}:
 \begin{eqnarray}
 \frac{dY_{\chi}}{dT} \approx -10^7 \mathrm{GeV}^3(\frac{m_2}{1 \mathrm{TeV}})^2(\frac{y_{sf}}{10^{-8}})^2 T^{-4}.
 \end{eqnarray}
  Thus $Y_\chi$ scales as $m_2^2\,y_{sf}^2$ for $T>m_2$. At $T\lesssim m_2$, the $h_2$ abundance becomes Boltzmann suppressed and $\chi$ production is no longer efficient. Therefore, we have:
 \begin{eqnarray}\label{y1}
 Y_{\chi}(T\lesssim m_2) \approx  10^{-4}(\frac{1\mathrm{TeV}}{m_2})(\frac{y_{sf}}{10^{-8}})^2,
 \end{eqnarray}
 The relic density of $\chi$, $\Omega_\chi h^2$, is related to the asymptotic value of $Y_\chi$ at low temperatures by:
 \begin{eqnarray}\label{om1}
 \Omega_{\chi}h^2 =2.744 \times 10^8 \frac{m_{\chi}}{\mathrm{GeV}}Y_{\chi}(T_0),
 \end{eqnarray}
 where $T_0=2.725$~K is the present-day cosmic microwave
background (CMB) temperature. For $\chi$ production via the freeze-in mechanism, the relic density can be estimated as \cite{Yaguna:2023kyu}:
\begin{eqnarray}
\Omega_{\chi}h^2 \approx 0.3 (\frac{m_{\chi}}{0.1\mathrm{GeV}})(\frac{1\mathrm{TeV}}{m_2})(\frac{y_{sf}}{10^{-10}})^2, 
\end{eqnarray}
 where we used Eq.~\ref{y1} and Eq.~\ref{om1}. For $m_2=1$~TeV, $m_\chi=1$~GeV, and assuming $\chi$ constitutes the entire DM relic density, one estimates $y_{sf}\lesssim\mathcal{O}(10^{-11})$. 
 
 For $m_\chi > m_2/2$, the decay $h_2 \to \chi\chi$ closes and $\chi$ production proceeds through
  SM-initiated channels $WW, h_1h_1, ZZ \to h_2^* \to \chi\chi$ as well as $h_2h_2 \to \chi\chi$. The SM-initiated contribution scales as $\Omega_{\rm SM}h^2\propto\sin^2\theta\cdot y_{sf}^2$, while the dark-sector process $h_2h_2\to\chi\chi$ contributes as $\Omega_{\rm Dark}h^2\propto y_{sf}^4$.
  For $m_\chi>m_2/2$, the allowed values of $y_{sf}$ satisfy $y_{sf}\lesssim 10^{-6}$ in the FIMP scenario, implying that SM-initiated processes can dominate $\chi$ production within the chosen parameter space.  Note that for smaller $\sin\theta$, e.g.\ $\sin\theta<10^{-6}$, $\chi$ production would be determined by the dark sector. Since the $2\to2$ process involves a two-body initial state, the threshold condition $\sqrt{s}\geq 2m_\chi$ involves the center-of-mass energy of the pair, not the energy of a single particle.  In the
  thermal bath, the kinetic energies of two  particles can combine to overcome the threshold, resulting in a much weaker suppression than the single-particle case, even though  this scattering becomes kinematically forbidden at zero temperature. Such ``forbidden freeze-in'' through scattering when the parent particle is lighter than the DM candidate has been systematically analyzed in Ref.~\cite{Li:2023ewv}.

  The production of $S$ and $\chi$ proceeds independently in the FIMP scenario. $S$ production receives contributions from both the visible sector ($h_1\to SS$ decay + $XX\to SS$ scattering) and the dark sector ($h_2\to SS$ decay + $h_2h_2\to SS$ scattering). The visible-sector contribution is consistent with the results of the singlet scalar FIMP model, and the $S$ relic density is determined by $m_S$, $m_2$, $\lambda_{ds}$, $\lambda_{dh}$ and $v_0=m_{\chi}/y_{sf}$. Note that although $\chi$ does not directly enter the $S$ production processes, $v_0$ appears in both $h_2h_2\to SS$ and $h_2\to SS$. For tiny $y_{sf}$, $v_0$ becomes very large, so that a correspondingly tiny $\lambda_{ds}$ is required to avoid overproduction of $S$. In other words, $\chi$ and $S$ are linked indirectly through $v_0$: the small $y_{sf}$ required for $\chi$ to achieve the correct relic abundance induces a large $v_0$, which in turn demands an extremely tiny $\lambda_{ds}$ to avoid overproducing $S$.
  
   \begin{figure}[htbp]
\centering
\includegraphics[height=7.5cm,width=17cm]{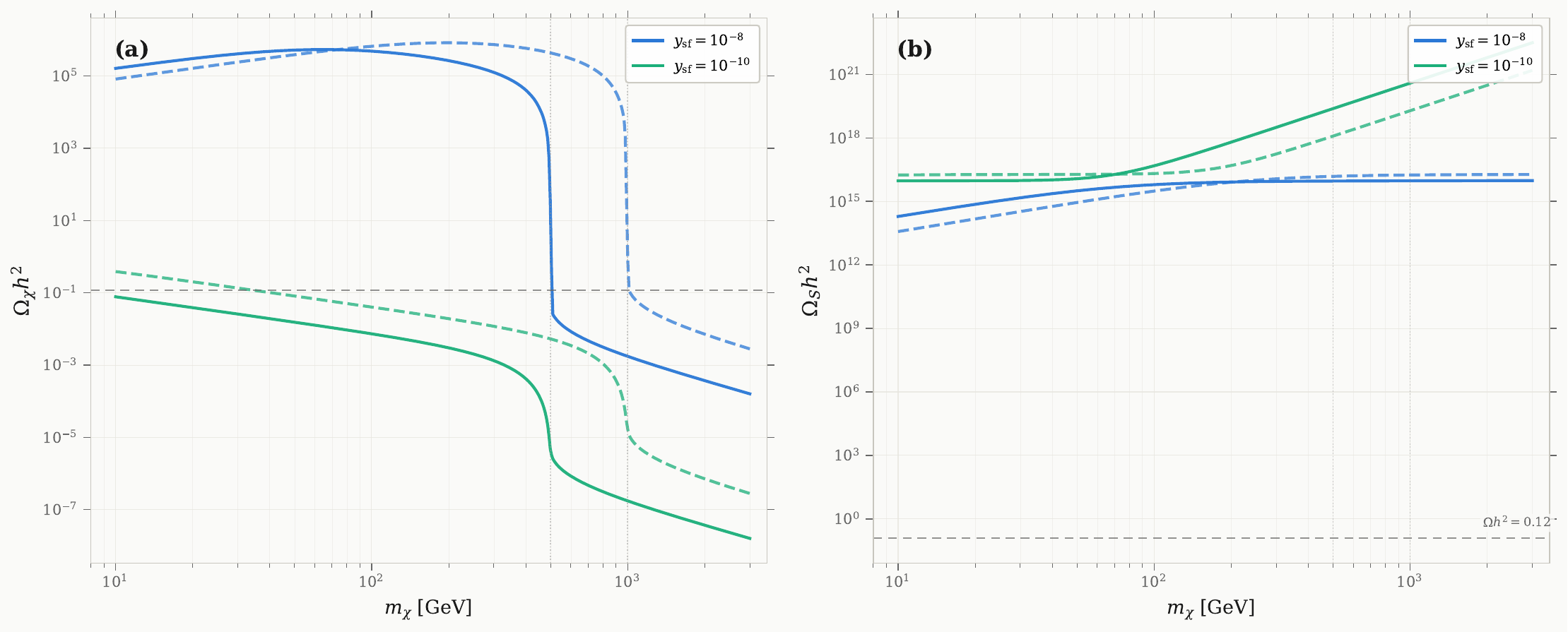}
\caption{ Evolution of $\Omega_\chi h^2$ (left) and $\Omega_S h^2$ (right) as functions of $m_\chi$, with $\lambda_{dh}=10^{-10}$, $\lambda_{ds}=10^{-10}$, $m_S=400$~GeV fixed. Note that $\lambda_{ds}=10^{-10}$ is chosen for illustration and exceeds the viable range found in Sec.~\ref{sec:discussion}. Different colors correspond to different values of $y_{sf}$; solid (dashed) lines correspond to $m_2=1$~TeV ($m_2=2$~TeV). The grey dashed line in the left panel indicates the observed DM relic density.}
\label{fig1}
\end{figure}
\begin{figure}[htbp]
\centering
\includegraphics[height=7.5cm,width=17cm]{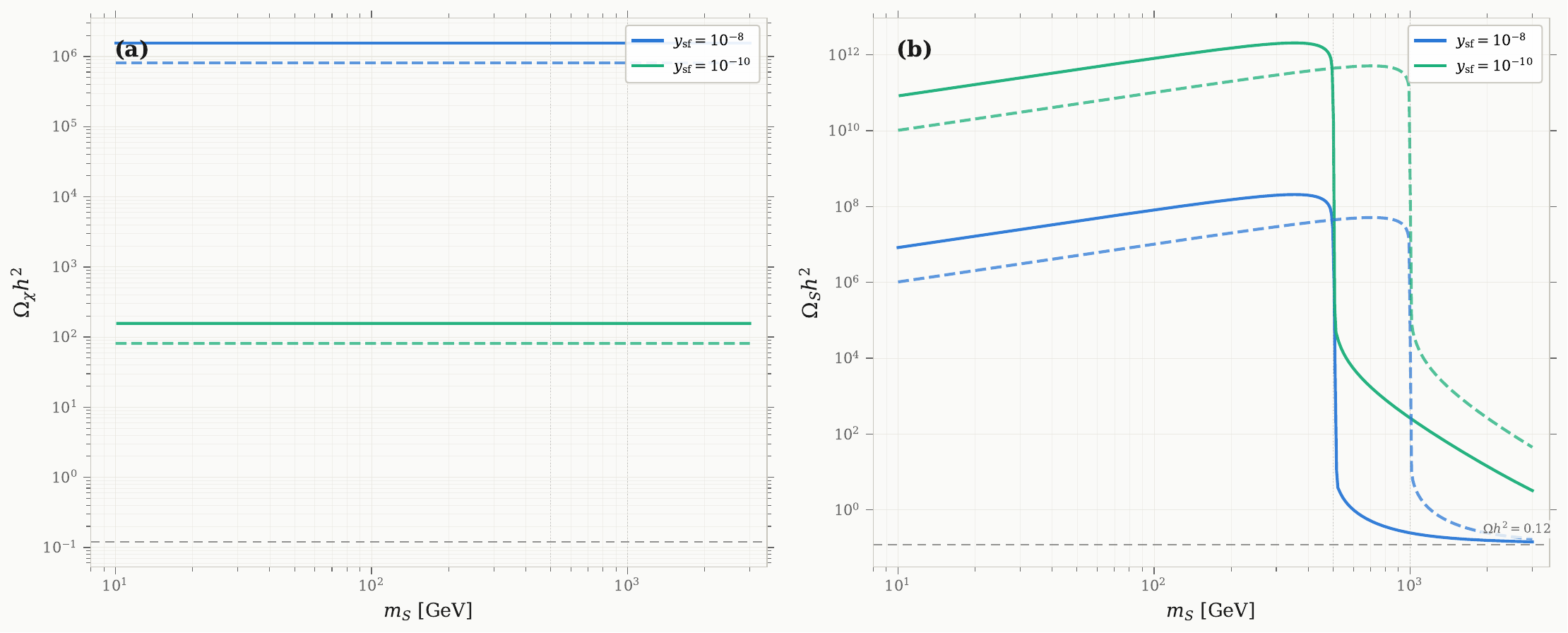}
\caption{ Evolution of $\Omega_\chi h^2$ (left) and $\Omega_S h^2$ (right) as functions of $m_S$, with $\lambda_{dh}=10^{-11}$, $\lambda_{ds}=10^{-14}$, $m_\chi=100$~GeV fixed. Different colors correspond to different values of $y_{sf}$; solid (dashed) lines correspond to $m_2=1$~TeV ($m_2=2$~TeV). The grey dashed line in the left panel indicates the observed DM relic density.}
\label{fig11}
\end{figure}

  In Fig.~\ref{fig1} we show $\Omega_\chi h^2$ and $\Omega_S h^2$ as functions of $m_\chi$, with $\lambda_{dh}=10^{-10}$, $\lambda_{ds}=10^{-10}$, $m_S=400$~GeV fixed.
In Fig.~\ref{fig1}(a), the behavior of $\Omega_\chi h^2$ can be divided into two regions: $m_\chi<m_2/2$, where decay dominates, and $m_\chi>m_2/2$, where scattering processes dominate. For $m_2=1$~TeV, when $m_\chi<500$~GeV, $\chi$ production is dominated by $h_2\to\chi\chi$ decay, and $\Omega_\chi h^2$ increases with $y_{sf}$ due to the larger decay rate. For fixed $y_{sf}$, $\Omega_\chi h^2$ increases with $m_\chi$, as seen in Fig.~\ref{fig1}(a). For $m_\chi>m_2/2$, $\chi$ production proceeds via $h_2h_2\to\chi\chi$ and $XX\to\chi\chi$, and the curve drops sharply for fixed $y_{sf}$. For $m_2=2$~TeV, the curves exhibit similar behavior, but drop sharply at $m_\chi\approx 1$~TeV.
In Fig.~\ref{fig1}(b), since $\lambda_{ds}=10^{-10}$ and the $h_2$-mediated $S$ production is enhanced by $v_0=m_\chi/y_{sf}$, the green lines, which correspond to a smaller $y_{sf}$, always lie above the blue ones for fixed $m_2$.

 In Fig.~\ref{fig11}, we show $\Omega_\chi h^2$ and $\Omega_S h^2$ as functions of $m_S$, with $\lambda_{dh}=10^{-11}$, $\lambda_{ds}=10^{-14}$, $m_\chi=100$~GeV fixed. The value of $\Omega_\chi h^2$ is almost unchanged as $m_S$ increases for fixed $m_\chi$ and $y_{sf}$, as can be seen in Fig.~\ref{fig11}(a).
 As for $\Omega_Sh^2$, one can see a sharp drop at around $m_S \approx m_2/2$ in Fig.~\ref{fig11}(b), where the decay $h_2\to SS$ becomes kinematically closed and $2\to2$ annihilation processes take over. For $m_S>m_2$, the scattering $h_2h_2\to SS$ itself becomes kinematically forbidden at zero temperature and proceeds only via the thermal tail, in complete analogy with the forbidden freeze-in scenario for $\chi$ discussed above~\cite{Li:2023ewv}. Consequently $\Omega_Sh^2$ decreases further with increasing $m_S$. We obtain a similar conclusion for $m_2=2$~TeV, where the curves drop at about $m_S=1$~TeV. Similarly, the green lines, which correspond to a smaller $y_{sf}$, lie above the blue ones for fixed $m_2$, as noted above.

We stress that the results shown in this and all the following figures were obtained with \textsc{micrOMEGAs}~\cite{Alguero:2023zol} and not with the analytical expressions obtained in the text, which serve instead as a check and illustrate the functional dependence on the different parameters.

  \section{Discussion}\label{sec:discussion}
 \subsection{Viable parameter space of the model for the FIMP regime}
 For $m_\chi<m_2/2$, the upper bound on $y_{sf}$ is approximately $y_{sf}<\mathcal{O}(10^{-11})$ for $m_2=1$~TeV; such a tiny value demands a correspondingly small $\lambda_{ds}$ to obtain the correct $\Omega_S h^2$. For $m_\chi>m_2/2$, $\chi$ production is determined by $2\to2$ processes and considerably larger values of $y_{sf}$ become viable, as can be seen from Fig.~\ref{fig1}. For $S$, the relic density increases with both portal couplings $\lambda_{dh}$ and $\lambda_{ds}$, and the relative contribution of the $h_2$-mediated channels versus the SM-mediated channels depends sensitively on whether $m_S$ lies below or above $m_2/2$.

We determine the viable parameter space by requiring the DM relic density to lie within $[0.11,\,0.13]$, corresponding to an approximately $10\%$ window around the Planck central value. This generous range accounts for theoretical uncertainties in the freeze-in computation, such as the temperature dependence of the effective number of degrees of freedom $g_{*\mathcal{S}}$. Moreover, we classify the model into four cases according to the mass hierarchy between the DM particles and $h_2$:
(i)~$m_{\chi}<m_2/2,\;m_S<m_2/2$;\quad
(ii)~$m_{\chi}<m_2/2,\;m_S>m_2/2$;\quad
(iii)~$m_{\chi}>m_2/2,\;m_S<m_2/2$;\quad
(iv)~$m_{\chi}>m_2/2,\;m_S>m_2/2$.

For definiteness, we fix $m_2=1$~TeV and perform a random scan over the following regions:
\begin{align}
\lambda_{dh} \in [10^{-14},\,10^{-11}],\qquad y_{sf} \in [10^{-14},\,10^{-6}],\qquad \lambda_{ds} \in [10^{-25},\,10^{-10}], \qquad m_{\chi,S} \in [1~\mathrm{GeV},3~\mathrm{TeV}].
\end{align}
The range adopted for $\lambda_{dh}$ is comparable to that of the standard singlet scalar FIMP model, whereas the lower bound of $\lambda_{ds}$ extends to values far below those encountered in the traditional scenarios.

\begin{figure}[htbp]
\centering
\includegraphics[height=6cm,width=18cm]{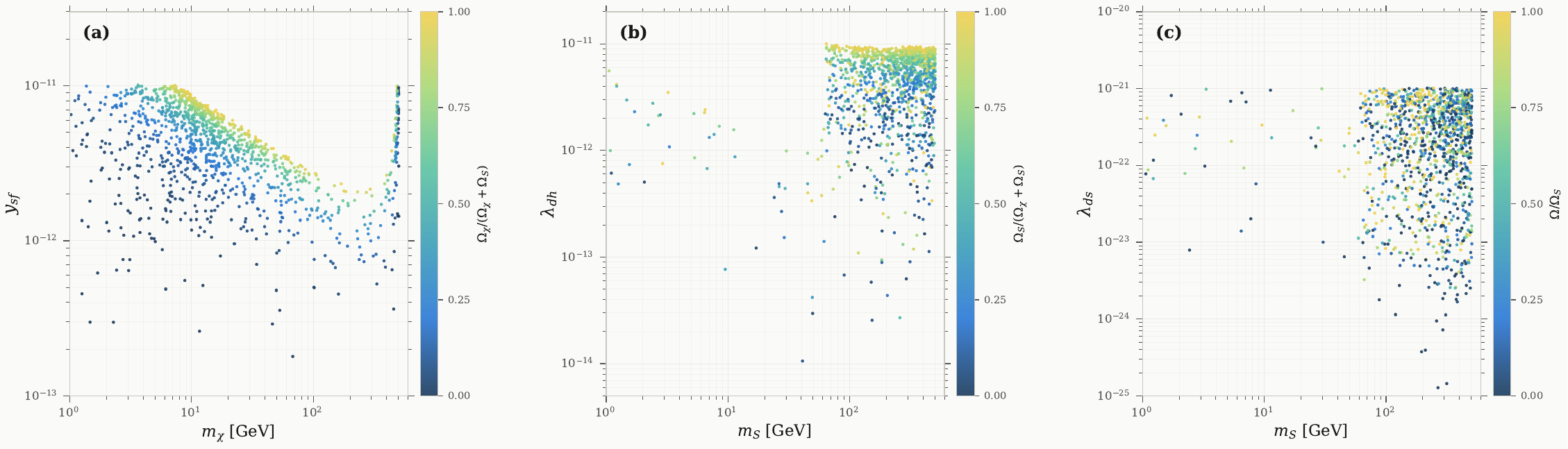}
\caption{Results for Case~(i): $m_{\chi}<500$~GeV and $m_S<500$~GeV. Panel~(a): viable parameter space in the $(m_{\chi},\,y_{sf})$ plane, with the color bar indicating the $\chi$ fraction $\Omega_{\chi}/(\Omega_S+\Omega_{\chi})$. Panel~(b): viable parameter space in the $(m_S,\,\lambda_{dh})$ plane, with the color bar indicating the $S$ fraction $\Omega_S/(\Omega_S+\Omega_{\chi})$. Panel~(c): viable parameter space in the $(m_S,\,\lambda_{ds})$ plane, with the color bar $\Omega/\Omega_S$ indicating the $h_2$ contribution to the $S$ relic density, here $\Omega$ denotes the yield from $h_2\to SS$.}\label{fig6}
\end{figure}

The results for $m_{\chi}<500$~GeV and $m_S<500$~GeV are presented in Fig.~\ref{fig6}. Here $\chi$ production proceeds mainly via the two-body decay $h_2\to\chi\chi$, while $S$ is produced through both the decay $h_2\to SS$ and $2\to2$ scattering $XX\to SS$, where $X$ denotes SM particles. Fig.~\ref{fig6}(a) displays the viable region of $(m_{\chi},\,y_{sf})$. The Yukawa coupling is constrained to $y_{sf}\in(3\times10^{-13},\,10^{-11}]$, and $m_{\chi}$ spans the full interval $[1~\mathrm{GeV},\,500~\mathrm{GeV}]$. For very light $\chi$ ($m_{\chi}\lesssim 7$~GeV), the decay width $\Gamma(h_2\to\chi\chi)\propto y_{sf}^2\,m_2$ is essentially independent of $m_{\chi}$, and $\chi$ remains a subdominant component of the total DM density regardless of $y_{sf}$. For $m_{\chi}\gtrsim 7$~GeV, a larger Yukawa coupling generically yields a larger $\chi$ fraction, and $\chi$ can become the dominant DM constituent for the highest allowed values of $y_{sf}$. As $m_{\chi}$ increases, the upper bound on $y_{sf}$ initially decreases because $\Omega_\chi h^2 \propto m_\chi$ for an approximately mass-independent decay width, so the over-abundance limit forces the allowed Yukawa coupling to decrease; however, for $m_{\chi}\gtrsim 300$~GeV, the phase-space suppression factor $(1-4m_\chi^2/m_2^2)^{3/2}$ reduces the decay rate, and the upper bound on $y_{sf}$ turns around and grows with $m_\chi$ in order to maintain the correct total relic density.

In Fig.~\ref{fig6}(b), we show the viable parameter space of $(m_S,\,\lambda_{dh})$, where the scalar DM mass $m_S$ covers the full interval $[1~\mathrm{GeV},\,500~\mathrm{GeV}]$, while $\lambda_{dh}$ spans the scanned range and most of the points lie in the upper-right region of the plane.  For $m_S<m_1/2$, the upper bound of the allowed value for $\lambda_{dh}$ decreases with the increase of $m_S$ under the DM relic density constraint, where the SM sector plays an important role in determining $S$ production. As $m_S$ increases beyond $m_1/2$, the viable range of $\lambda_{dh}$ broadens and becomes less constrained. For $\lambda_{dh}$ as small as $8\times10^{-14}$, $S$ can still constitute the dominant DM component, provided the $h_2$-mediated channels play a dominant role in determining the $S$ relic density. Figure~\ref{fig6}(c) depicts the $(m_S,\,\lambda_{ds})$ parameter space, where points with different colors represent the fractional contribution of $h_2\to SS$ to $\Omega_S h^2$, denoted by $\Omega/\Omega_S$. The portal coupling $\lambda_{ds}$ is constrained within $(10^{-25},\,10^{-21}]$, values that are far smaller than those encountered in traditional FIMP models, and most of the viable points are concentrated in the upper-right part of the plane with $\Omega/\Omega_S$ spanning the interval $(0,1)$.

\begin{figure}[htbp]
\centering
\includegraphics[height=6cm,width=18cm]{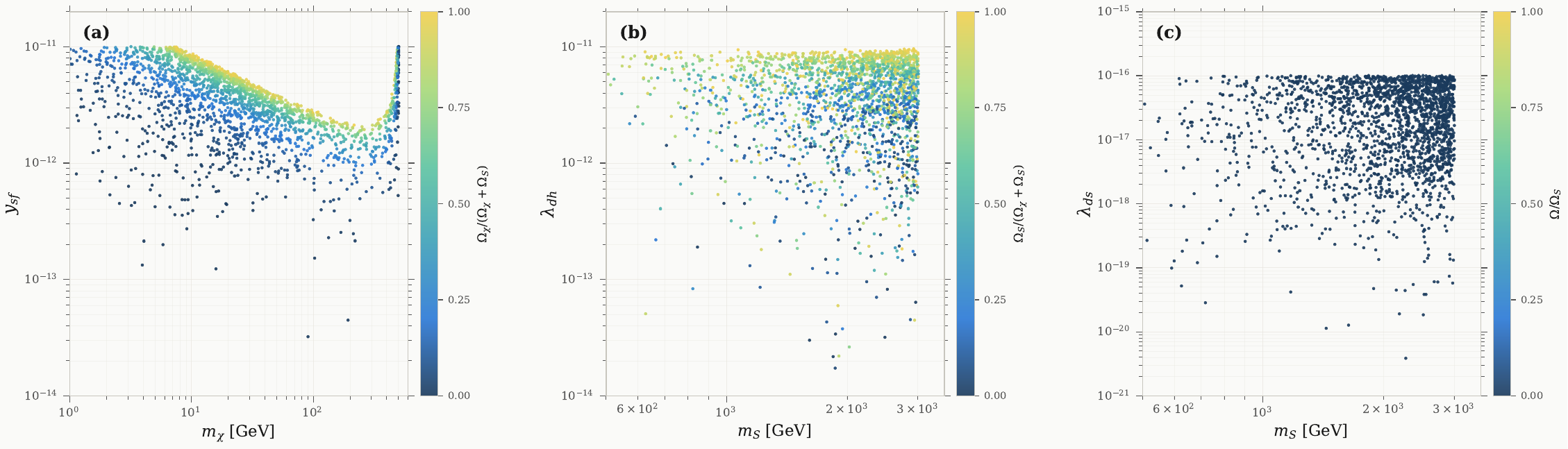}
\caption{Same as Fig.~\ref{fig6}, but for Case~(ii): $m_{\chi}<500$~GeV and $500~\mathrm{GeV}<m_S<3000$~GeV, and $\Omega$ in Panel~(c) denotes the yield from $h_2 h_2\to SS$.}\label{fig7}
\end{figure}

We display the results for $m_{\chi}<500$~GeV and $m_S>500$~GeV in Fig.~\ref{fig7}. Since the $\chi$ mass range is unchanged with respect to Case~(i), $\chi$ production is still mainly governed by $h_2\to\chi\chi$, and the $(m_{\chi},\,y_{sf})$ parameter space shown in Fig.~\ref{fig7}(a) exhibits the same qualitative features already described for Fig.~\ref{fig6}(a). The $(m_S,\,\lambda_{dh})$ parameter space is presented in Fig.~\ref{fig7}(b), and the scalar DM mass $m_S$ covers the full interval $(500~\mathrm{GeV},\,3~\mathrm{TeV}]$, while $\lambda_{dh}$ spans the scanned range and most of the points lie in the upper-right region of the plane.
For larger values of $\lambda_{dh}$, the SM-mediated $XX\to SS$ processes become efficient, and $S$ can dominate the total DM relic density as can be seen in Fig.~\ref{fig7}(c), where the contribution of $h_2$ to $S$ production is much smaller. Moreover, according to Fig.~\ref{fig7}(c),
the coupling $\lambda_{ds}$ is bounded within $(4\times10^{-21},\,10^{-16}]$, which is also smaller than in traditional FIMP models, as shown in Case~(i).

\begin{figure}[htbp]
\centering
\includegraphics[height=6cm,width=18cm]{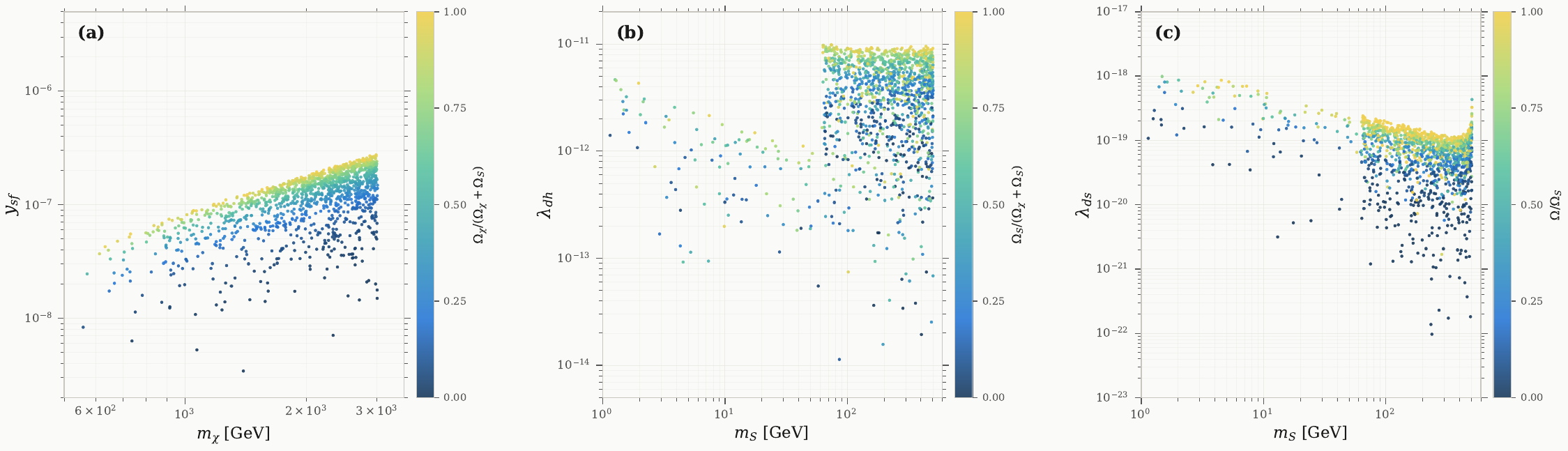}
\caption{Same as Fig.~\ref{fig6}, but for Case~(iii): $500~\mathrm{GeV}<m_{\chi}<3$~TeV and $m_S<500$~GeV, and $\Omega$ in Panel~(c) denotes the yield from $h_2\to SS$.}
\label{fig8}
\end{figure}

In Fig.~\ref{fig8}, we show the results for $m_{\chi}>500$~GeV and $m_S<500$~GeV. In this region, the decay $h_2\to\chi\chi$ is kinematically closed, and $\chi$ production is mediated by the $2\to2$ annihilations. According to Fig.~\ref{fig8}(a), $y_{sf}$ is now constrained to $(2\times10^{-9},\,3\times10^{-7})$, while $m_{\chi}$ spans the entire interval $[500~\mathrm{GeV},\,3~\mathrm{TeV}]$. For fixed $m_{\chi}$, a larger $y_{sf}$ enhances the $h_2h_2\to\chi\chi$ cross section and thereby increases the $\chi$ fraction, and 
for $y_{sf}\gtrsim 2 \times 10^{-7}$, $\chi$ always constitutes the dominant DM component. On the other hand, with the increase of $m_{\chi}$, the upper bound of the viable $y_{sf}$ value increases to obtain the correct DM relic density.
The allowed parameter space for $(m_S,\,\lambda_{dh})$ in Fig.~\ref{fig8}(b) is qualitatively similar to that of Case~(i), since $m_S<m_2/2$ and $S$ production is still determined by $h_2\to SS$ together with $XX\to SS$ scattering. We show  $(m_S,\,\lambda_{ds})$ parameter space in Fig.~\ref{fig8}(c) where $\lambda_{ds}$ lies in the range $(10^{-22},\,10^{-18})$.

For fixed $m_S$,  contribution of $h_2$ to $S$ relic density will be more efficient with a larger $\lambda_{ds}$ as $S$ is light, and the process $h_2 \to SS$ can contribute to the dominant $S$  constituent for the highest allowed values of $\lambda_{ds}$. As $m_S$ increases, the upper bound on $\lambda_{ds}$ initially decreases because $\Omega_S h^2 \propto m_S$ for an approximately mass-independent decay width, so the over-abundance limit forces the allowed $\lambda_{ds}$ to decrease; however, for $m_{S}\gtrsim 400$~GeV, the phase-space suppression reduces the decay rate, and the upper bound on $\lambda_{ds}$ turns around and grows with $m_S$ to maintain the correct total relic density. The behavior of $\lambda_{ds}$ with $m_S$ is similar to that of $y_{sf}$ with $m_\chi$ in Case~(i), as discussed above.

\begin{figure}[htbp]
\centering
\includegraphics[height=6cm,width=18cm]{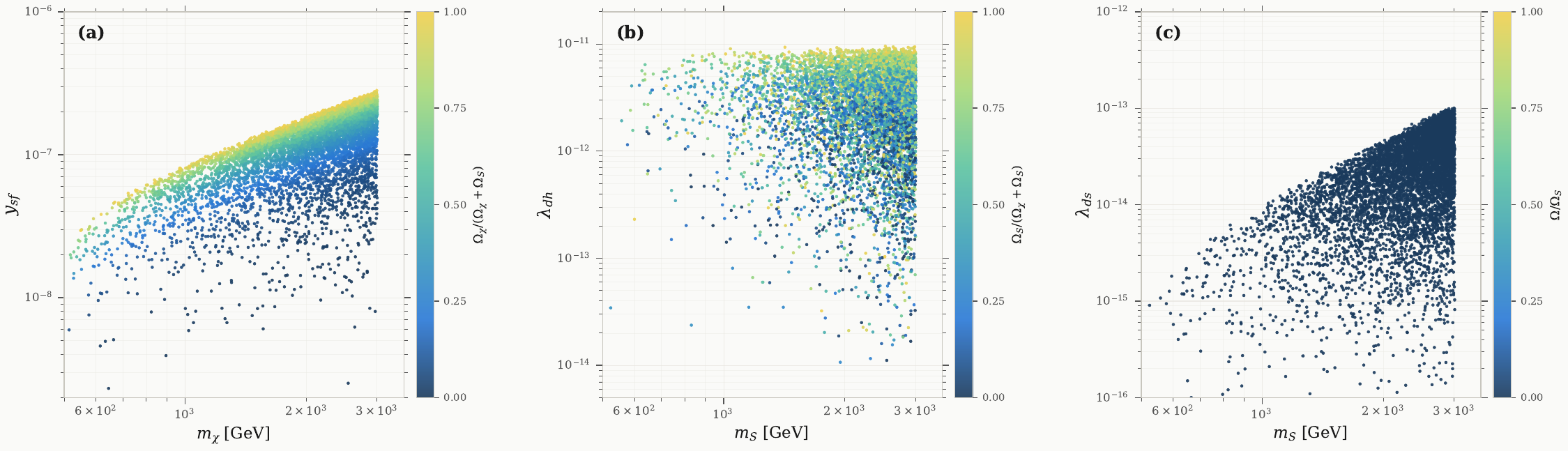}
\caption{Same as Fig.~\ref{fig6}, but for Case~(iv): $500~\mathrm{GeV}<m_{\chi}<3$~TeV and $500~\mathrm{GeV}<m_S<3$~TeV, and  $\Omega$ in Panel~(c) denotes the yield from $h_2  h_2\to SS$.}
\label{fig9}
\end{figure}

We present the results for $m_{\chi}>500$~GeV and $m_S>500$~GeV in Fig.~\ref{fig9}, where both DM particles are heavier than $m_2/2$ and all production proceeds through $2\to2$ annihilation processes.
Fig.~\ref{fig9}(a) shows that the Yukawa coupling is bounded within $(10^{-9},\,3\times10^{-7}]$, while $m_{\chi}$ can take any value in $[500~\mathrm{GeV},\,3~\mathrm{TeV}]$. As in Case~(iii), a larger $y_{sf}$ always yields a larger $\chi$ fraction, with $\chi$ becoming the dominant component for the highest allowed couplings. The parameter space for $(m_S,\,\lambda_{dh})$  is given in Fig.~\ref{fig9}(b), and the scalar DM mass $m_S$ covers the full interval $(500~\mathrm{GeV},\,3~\mathrm{TeV}]$, while $\lambda_{dh}$ spans the scanned range as in Case~(ii). We display the result of $(m_S,\,\lambda_{ds})$ in Fig.~\ref{fig9}(c), where $\Omega$  corresponds to the yield from $h_2h_2\to SS$. The coupling $\lambda_{ds}$ is restricted to $[10^{-16},\,10^{-13}]$, and the $h_2$-mediated channel is always subdominant ($\Omega/\Omega_S\ll 1$) regardless of $\lambda_{ds}$ value. On the other hand, with the increase of $m_S$, the upper bound of the allowed $\lambda_{ds}$ increases to obtain the correct DM relic density result.

In summary, both $m_S$ and $m_{\chi}$ are viable over the wide range $[1~\mathrm{GeV},\,3~\mathrm{TeV}]$ under the observed DM relic density constraint. The dominant production channels, and consequently the allowed coupling ranges, depend sensitively on the mass hierarchy between the DM particles and $h_2$. 
For $m_{\chi}<m_2/2$, $\chi$ production is governed by the decay $h_2\to\chi\chi$ with $y_{sf}\sim 10^{-13}$--$10^{-11}$, whereas for $m_{\chi}>m_2/2$, the $2\to2$ channels  take over and require $y_{sf}\sim 10^{-9}$--$10^{-7}$. The Higgs-portal coupling $\lambda_{dh}$ admits a comparatively flexible parameter space across all four cases, since its effect can always be compensated by adjusting the other free parameters. 
By contrast, $\lambda_{ds}$ exhibits four distinct allowed windows, one for each mass hierarchy, reflecting its interplay with the $Z_4$ breaking scale $v_0=m_{\chi}/y_{sf}$: when $y_{sf}$ is extremely small, $v_0$ becomes very large, which enhances the $h_2$-mediated $S$ production cross section and constrains $\lambda_{ds}$ to values much smaller than those in traditional FIMP models. When $m_S>m_2/2$, the contribution of $h_2$ to $S$ production is
highly suppressed due to the heavy mediator regardless of $\lambda_{ds}$, as shown in Fig.~\ref{fig7}(c) and Fig.~\ref{fig9}(c). When $m_S<m_2/2$,
even for such tiny values of $\lambda_{ds}$, the $h_2$-related processes can still constitute the dominant source of the $S$ relic density as shown in Fig.~\ref{fig6}(c) and Fig.~\ref{fig8}(c).

We close this discussion with a brief comment on the $m_2$ dependence of the above results. While the full parameter scan has been performed at $m_2=1$~TeV, the scaling of the viable coupling windows with $m_2$ can be understood analytically, as partially illustrated by the $m_2=2$~TeV curves in Figs.~\ref{fig1} and~\ref{fig11}.

For $\chi$ with $m_\chi<m_2/2$, the relic density scales as $\Omega_\chi h^2\propto m_\chi\,y_{sf}^2/m_2$~\cite{Yaguna:2023kyu}; thus, for a heavier $h_2$, a proportionally larger $y_{sf}$ is required to obtain the same relic abundance, and the allowed Yukawa window shifts upward roughly as $\sqrt{m_2}$. For $m_\chi>m_2/2$, both SM-initiated scattering ($XX\to h_2^*\to\chi\chi$) and dark-sector scattering ($h_2h_2\to\chi\chi$) contribute. The $m_2$ dependence of the SM channel is non-trivial---it ranges from $\propto 1/m_2^4$ in the low-temperature tail to approximately $m_2$-independent when $T\gg m_2$---while the dark-sector channel scales as $y_{sf}^4/m_2$. The net effect is that the viable $y_{sf}$ range shifts with $m_2$, but the hierarchical gap between the decay-dominated and scattering-dominated coupling windows (spanning several orders of magnitude) is sufficiently large that the qualitative structure of the parameter space is preserved across the range $m_2\in[300~\mathrm{GeV},3~\mathrm{TeV}]$.

For $S$ in the decay regime ($m_S<m_2/2$), where the $\lambda_{ds}v_0$ term dominates $h_2\to SS$, the scaling is particularly clean: $\Omega_S h^2\propto m_S\,\lambda_{ds}^2\,v_0^2/m_2^3$, so the $\lambda_{ds}$ window shifts as $m_2^{3/2}$. The most important qualitative effect of varying $m_2$, however, is the shift of the kinematic threshold $m_2/2$ that separates the decay-dominated and scattering-dominated regimes.

Regarding the sensitivity to $\sin\theta$, the arguments presented in Sec.~\ref{sec:sintheta} apply equally to the full analysis: for $m_{\chi,S}<m_2/2$, the decay widths that dominate DM production are $\sin\theta$-independent at leading order (since $\cos\theta\approx 1$), while for $m_{\chi,S}>m_2/2$, the SM-initiated scattering channels scale as $\sin^2\theta$ and can be compensated by a modest rescaling of $y_{sf}$ or $\lambda_{ds}$. Varying $\sin\theta$ within the experimentally allowed range ~\cite{Lopez-Val:2014jva,Robens:2021rkl} therefore modifies the precise numerical boundaries of the viable coupling windows but preserves the qualitative four-case classification and the hierarchical gap between decay- and scattering-dominated regimes. The values $\sin\theta=10^{-4}$ and $m_2=1$~TeV should thus be regarded as representative benchmarks that capture all the essential physics of the model.

\subsection{Naturalness considerations}\label{sec:naturalness}

The viable parameter space identified above involves two features that merit a discussion of their theoretical consistency: the large hierarchy between the symmetry-breaking scale $v_0$ and the electroweak scale, and the extremely small values of the portal coupling $\lambda_{ds}$. We show below that both features are radiatively stable and consistent with known bounds.

\subsubsection*{Hierarchy of the $Z_4$ breaking scale}

The fermion mass is generated through $m_\chi = y_{sf} v_0$, where $v_0$ is the vacuum expectation value of $S_0$. As discussed above, the viable parameter regions with $m_\chi$ at the GeV--TeV scale correspond to Yukawa couplings $y_{sf}\sim 10^{-9}$--$10^{-7}$ for $m_\chi>m_2/2$ and $y_{sf}\sim 10^{-13}$--$10^{-11}$ for $m_\chi<m_2/2$. Consequently, the symmetry-breaking scale
\begin{equation}
v_0 = \frac{m_\chi}{y_{sf}}
\end{equation}
typically lies in the ranges $[10^{9}$--$10^{12}]$~GeV and $[10^{11}$--$10^{15}]$~GeV respectively. Such a large hierarchy between $v_0$ and the electroweak scale $v=246$~GeV is a generic consequence of the FIMP scenario and raises the question of radiative stability in the scalar sector.

We note that a comparable hierarchy is not uncommon in freeze-in models. For instance, in the minimal fermion FIMP model~\cite{Yaguna:2023kyu}, the required Yukawa couplings are similarly tiny, and the associated new physics scale can be as high as the GUT scale. At the opposite extreme, freeze-in at stronger coupling~\cite{Cosme:2023fsc} shows that when the reheating temperature is below the DM mass, the required portal coupling can be as large as $\mathcal{O}(1)$, highlighting the breadth of the freeze-in parameter space. From a bottom-up perspective, the smallness of $y_{sf}$ is technically natural in the sense of 't~Hooft: in the limit $y_{sf}\to 0$, the Lagrangian acquires an enhanced chiral symmetry for $\chi$, and the beta function of $y_{sf}$ is proportional to $y_{sf}$ itself. The fermion sector hierarchy is therefore radiatively stable.

On the other hand, the large $v_0$ could potentially destabilize the electroweak scale through the Higgs portal coupling $\lambda_{sh}S_0^2|H|^2$. After $S_0$ acquires its vev, this term contributes $\lambda_{sh}v_0^2$ to the Higgs mass parameter. Using Eq.~\eqref{para_quartic}, $\lambda_{sh} = \sin2\theta\,(m_2^2-m_1^2)/(2vv_0)$, one finds
\begin{equation}
\lambda_{sh}v_0^2 = \frac{\sin2\theta\,(m_2^2-m_1^2)}{2v}\,v_0 \approx (4\times 10^{14}\,\mathrm{GeV}^2)\left(\frac{\sin\theta}{10^{-4}}\right)\left(\frac{v_0}{10^{15}\,\mathrm{GeV}}\right)\left(\frac{m_2}{1\,\mathrm{TeV}}\right)^2,
\end{equation}
which, for $v_0\sim 10^{15}$~GeV, exceeds the physical Higgs mass squared $m_1^2\approx 1.6\times10^4$~GeV$^2$ by some ten orders of magnitude. This contribution is, however, a tree-level tadpole that is absorbed into the minimization condition of the full scalar potential; it determines the mutual arrangement of $v$ and $v_0$ and does not represent a radiative correction to the Higgs mass. The physically relevant question is whether quantum corrections proportional to $v_0$ destabilize the electroweak scale. The leading one-loop correction to $m_1^2$ arises from $h_2$--$h_1$ mixing and scales as
\begin{equation}
\delta m_1^2 \sim \frac{\lambda_{sh}}{16\pi^2}\,m_2^2\log\!\left(\frac{\Lambda^2}{m_2^2}\right) \sim \frac{\sin2\theta\,m_2^4}{32\pi^2\,v\,v_0}\log\!\left(\frac{\Lambda^2}{m_2^2}\right),
\end{equation}
which is doubly suppressed---by the loop factor and by $1/v_0$---and amounts to $\delta m_1^2\sim 10^{-10}$--$10^{-4}$~GeV$^2$ for $v_0\sim 10^{15}$--$10^{9}$~GeV, entirely negligible compared to $m_1^2$. The hierarchy $v_0\gg v$ is therefore technically natural: in the decoupling limit $v_0\to\infty$ (equivalently $\lambda_{sh}\to 0$ with $m_2$ fixed), the two scalar sectors decouple, all radiative corrections from the heavy sector vanish, and the electroweak scale is protected by the enhanced symmetry of the decoupled theory.

\subsubsection*{Radiative stability of $\lambda_{ds}$}

The viable parameter space contains values of $\lambda_{ds}$ as small as $\sim\!10^{-25}$--$10^{-21}$ (for $m_\chi<m_2/2,\ m_S<m_2/2$). However, the model simultaneously contains the couplings $\lambda_{dh}S^2|H|^2$ and $\lambda_{sh}S_0^2|H|^2$. Even if $\lambda_{ds}$ is set to a tiny value at tree level, it will be regenerated radiatively through Higgs-mediated loop diagrams. In the unbroken phase, the full Higgs doublet $H$ runs in the loop, connecting the two portal couplings. The dominant additive  contribution can be obtained from the one-loop renormalization-group evolution of $\lambda_{ds}$:
\begin{equation}
\delta\lambda_{ds} \sim \frac{\lambda_{dh}\lambda_{sh}}{16\pi^2}\log\left(\frac{\Lambda^2}{m_{1,2}^2}\right),
\end{equation}
where $\Lambda$ is the renormalization scale. Using the relation $\lambda_{sh} = \sin2\theta(m_2^2-m_1^2)/(2vv_0)$ from Eq.~\eqref{para_quartic}, and taking $\cos\theta\approx 1$ for $\sin\theta\ll 1$, one obtains:
\begin{equation}
\delta\lambda_{ds} \approx \frac{\lambda_{dh}\sin\theta\,(m_2^2-m_1^2)}{8\pi^2\,v\,v_0}\log\left(\frac{\Lambda}{m_2}\right) \approx 5.1\times 10^{-24}\left(\frac{\lambda_{dh}}{10^{-12}}\right)\left(\frac{\sin\theta}{10^{-4}}\right)\left(\frac{10^{10}\,\mathrm{GeV}}{v_0}\right)\left(\frac{\log(\Lambda/m_2)}{10}\right)\left(\frac{m_2}{1\,\mathrm{TeV}}\right)^{\!2}.
\end{equation}

For the parameter ranges considered in this work, $\delta\lambda_{ds}$ is below the tree-level values of $\lambda_{ds}$ shown in Figs.~\ref{fig6}--\ref{fig9}, and the radiative stability is reinforced by the correlation between $v_0$ and $\lambda_{ds}$: smaller $\lambda_{ds}$ requires larger $v_0$, which in turn suppresses $\delta\lambda_{ds}\propto 1/v_0$. Taking the most aggressive parameter choices---$\lambda_{dh}\sim 10^{-11}$, $\sin\theta\sim 10^{-4}$, $v_0\sim 10^{9}$~GeV, and $\log(\Lambda/m_2)\sim 10$---one obtains $\delta\lambda_{ds}\sim 5.1\times 10^{-22}$. While this approaches the lower edge of the tree-level $\lambda_{ds}$ range in the corresponding regime (Case~(iii), where $\lambda_{ds}\in[10^{-22},10^{-18}]$), such extreme choices (maximal $\lambda_{dh}$, minimal $v_0$) are typically realized for parameter points in the upper portion of the $\lambda_{ds}$ window, where the tree-level coupling is larger. In the regime where $\lambda_{ds}$ reaches its smallest values ($m_\chi<m_2/2$, $m_S<m_2/2$), $v_0\gtrsim 10^{11}$~GeV and the correction is safely suppressed. For instance, with $v_0\sim 10^{13}$~GeV (corresponding to $y_{sf}\sim 10^{-11}$, $m_\chi\sim 100$~GeV), $\lambda_{dh}\sim 10^{-12}$, and $\sin\theta\sim 10^{-4}$, one finds $\delta\lambda_{ds}\sim 5.1\times 10^{-27}$, over six orders of magnitude below the required $\lambda_{ds}\sim 10^{-20}$. We therefore conclude that the tiny values of $\lambda_{ds}$ required by the FIMP scenario are technically natural and stable under radiative corrections.

\subsubsection*{Gravitational floor for $\lambda_{ds}$}

An intriguing consequence of the extreme smallness of $\lambda_{ds}$ is the question of whether there exists a fundamental lower bound beyond which the portal coupling becomes physically irrelevant. Such a bound is provided by gravitational freeze-in~\cite{Garny:2018ali,Mambrini:2021gpdm,Bernal:2021gdp}: even in the complete absence of non-gravitational interactions, DM can be produced through graviton-mediated scattering $XX\to SS$ (where $X$ denotes any particle in thermal equilibrium). The gravitational production rate per unit volume scales as $\gamma_{\rm grav}\sim T^8/M_{\rm Pl}^4$, where $M_{\rm Pl}=1.22\times10^{19}$~GeV. Crucially, this rate grows steeply with temperature and its contribution to the DM relic density is dominated by the highest temperature reached by the thermal bath, i.e.\ the reheating temperature $T_{\rm RH}$. The $\lambda_{ds}$-mediated production, by contrast, is dominated by $T\sim m_2$ where the $h_2$ abundance peaks, and is essentially independent of $T_{\rm RH}$ (provided $T_{\rm RH}\gtrsim m_2$, as required for $h_2$ thermalization).

Integrating the Boltzmann equations for the $S$ abundance, the relic yield $Y_S\equiv n_S/s$ receives a contribution from the $\lambda_{ds}$-mediated channel that scales as $Y_S|_{\lambda_{ds}}\propto\lambda_{ds}^2 v_0^2 M_{\rm Pl}/m_2^3$, while gravitational production contributes $Y_{\rm grav}\propto T_{\rm RH}^3/M_{\rm Pl}^3$~\cite{Garny:2018ali,Kolb:2023cgpp}. Equating the two yields the \textit{gravitational floor}:
\begin{equation}\label{eq:gravfloor}
\lambda_{ds}^{\rm floor}\simeq \frac{m_2^{3/2}\,T_{\rm RH}^{3/2}}{v_0\,M_{\rm Pl}^2}
= 7\times 10^{-43}\left(\frac{m_2}{1~{\rm TeV}}\right)^{\!3/2}
\left(\frac{T_{\rm RH}}{1~{\rm TeV}}\right)^{\!3/2}
\left(\frac{10^{13}~{\rm GeV}}{v_0}\right).
\end{equation}
For the minimal reheating temperature $T_{\rm RH}\sim m_2\sim 1$~TeV and $v_0$ in the range $[10^9,\,10^{15}]$~GeV, $\lambda_{ds}^{\rm floor}$ spans $[7\times10^{-45},\,7\times10^{-39}]$, lying 15--20 orders of magnitude below the smallest $\lambda_{ds}$ values ($\sim10^{-25}$) encountered in our scan. Even for the most extreme case---$T_{\rm RH}\sim M_{\rm Pl}$ and $v_0\sim10^9$~GeV, corresponding to the upper edge of the scattering-dominated regime---one finds $\lambda_{ds}^{\rm floor}\sim 10^{-14}$, still an order of magnitude below the largest viable $\lambda_{ds}$ in that regime ($\sim10^{-13}$). The gravitational floor therefore does not constrain the viable parameter space; rather, it demonstrates that the FIMP production of $S$ via the $h_2$ portal remains the dominant mechanism for $\lambda_{ds}$ values all the way down to $10^{-39}$--$10^{-45}$ for natural reheating scenarios. Below this floor, $\lambda_{ds}$ loses its physical meaning as gravitational production takes over, providing an irreducible contribution assuming standard radiation-dominated reheating.

\subsubsection*{UV interpretation of hierarchically small $\lambda_{ds}$}

The values $\lambda_{ds}\sim10^{-25}$--$10^{-13}$ obtained in our scan are far smaller than any known Standard Model Yukawa coupling. While the preceding radiative stability analysis establishes that these values are not destabilized by quantum corrections, it says nothing about their dynamical origin. A compelling UV interpretation would explain \emph{why} $\lambda_{ds}$ is so small, rather than merely accommodating it. Several well-known mechanisms can naturally generate hierarchically small dimensionless couplings:

\medskip\noindent\textit{(i)~Higher-dimensional operator.} If a UV symmetry (e.g.\ an extended discrete group $Z_N\times Z_M$ with $N,M>4$) forbids the renormalizable portal $S^2S_0^2$, the lowest allowed operator may be of dimension six or higher. For instance, $S^2 S_0^2 |H|^2/\Lambda^2$ generates, after electroweak symmetry breaking, an effective $\lambda_{ds}^{\rm eff}\sim v^2/(2\Lambda^2)$. To obtain $\lambda_{ds}\sim10^{-25}$--$10^{-13}$ requires $\Lambda\sim 5\times10^{8}$--$5\times10^{14}$~GeV, a range that encompasses the seesaw scale, the Peccei--Quinn scale, and the GUT scale---scales already well-motivated by independent considerations.

\medskip\noindent\textit{(ii)~Radiative generation.} If $\lambda_{ds}=0$ at tree level due to a symmetry, it can be generated at loop level: $\lambda_{ds}\sim(g^2/16\pi^2)^n$, where $g$ is a typical coupling and $n$ the loop order. For $g\sim0.3$, a six- to eight-loop ($n=6$--$8$) suppression yields $\lambda_{ds}\sim10^{-20}$--$10^{-26}$, comfortably within our viable range. For $g\sim0.1$, $n=4$--$6$ covers $\lambda_{ds}\sim10^{-17}$--$10^{-26}$. This mechanism requires no high scale---only an accidental or imposed symmetry that forbids the tree-level portal.

\medskip\noindent\textit{(iii)~Froggatt--Nielsen mechanism~\cite{Froggatt:1978nt}.} A horizontal $U(1)_{\rm FN}$ symmetry, spontaneously broken by a flavon field $\phi$ with $\langle\phi\rangle/\Lambda_{\rm FN}=\varepsilon$, assigns different charges to $S^2$ and $S_0^2$, so that $\lambda_{ds}\propto\varepsilon^{|q|}$. With the canonical value $\varepsilon\sim0.2$ (the Cabibbo angle), charge differences $|q|=18$--$36$ produce $\lambda_{ds}\sim10^{-13}$--$10^{-25}$, naturally covering our entire $\lambda_{ds}$ window. Smaller $\varepsilon$ values, common in the lepton sector of FN constructions, require correspondingly fewer charge units.

\medskip
Each of these mechanisms replaces the apparent fine-tuning of a free parameter with structural features---symmetries or flavor physics---that are independently studied and tested in other arenas. The tiny $\lambda_{ds}$ is therefore not an embarrassment for the model but a potential window into its UV completion. While a detailed construction lies beyond the scope of this work, the existence of multiple viable UV scenarios supports the theoretical consistency of the FIMP-FIMP regime across the full $\lambda_{ds}$ range identified in our scan.

We close the discussion with a brief note on cosmological consistency. The non-thermal momentum distribution of FIMP DM can, in principle, lead to free-streaming that erases small-scale structure; however, for the GeV--TeV DM masses considered here, the free-streaming length is $\lambda_{\rm fs}\sim10^{-7}$--$10^{-10}$~Mpc~\cite{Choi:2023plb}, corresponding to an effective thermal warm DM mass $m_{\rm WDM}^{\rm eff}\gtrsim10^4$~keV---over three orders of magnitude above the most stringent Lyman-$\alpha$ bound of $5.7$~keV~\cite{Irsic:2024lya}. Both $S$ and $\chi$ are therefore firmly in the cold DM regime.

\subsection{Comparison with WIMP-WIMP and mixed WIMP-FIMP regimes}
\begin{table}[htbp]
\centering
\caption{Comparison of the three dark-sector regimes realized within the same $Z_2\times Z_4$ model. For the mixed WIMP-FIMP regime~\cite{Qi:2025znq}, the two sub-cases are shown separately.}
\label{tab:comparison}
\begin{tabular}{lcccc}
\toprule
 & WIMP-WIMP~\cite{Qi:2025jpm} & \makecell{Mixed (Case~I)\\\cite{Qi:2025znq}} & \makecell{Mixed (Case~II)\\\cite{Qi:2025znq}} & FIMP-FIMP (this work) \\
\midrule
$\chi$ production   & freeze-out  & freeze-out   & freeze-in    & freeze-in \\
$S$ production      & freeze-out  & freeze-in    & freeze-out   & freeze-in \\
$y_{sf}$            & $0.02$--$3.14$  & $\gtrsim 1$ & $10^{-13}$--$10^{-7}$ & $10^{-13}$--$10^{-7}$ \\
$\lambda_{ds}$      & $10^{-5}$--$3.14$ & $10^{-15}$--$10^{-7}$ & $10^{-13}$--$10^{-7}$ & $10^{-25}$--$10^{-13}$ \\
$\lambda_{dh}$      & $\gtrsim 0.2$ & $10^{-15}$--$10^{-7}$ & $\sim 10^{-1}$ & $10^{-14}$--$10^{-11}$ \\
Direct detection    & accessible & $\chi$ accessible & $S$ accessible   & both invisible \\
Indirect detection  & accessible & $\chi$ accessible & $S$ accessible   & both invisible \\
\bottomrule
\end{tabular}
\end{table}

The same $Z_2\times Z_4$ Lagrangian admits three qualitatively distinct dark-sector scenarios depending on the production mechanism of each DM species, completing a consistent phenomenological picture without any modification of the particle content or symmetries. Table~\ref{tab:comparison} summarizes the key features of the three regimes. The mixed WIMP-FIMP scenario~\cite{Qi:2025znq} comprises two sub-cases---Case~I ($\chi$ as WIMP, $S$ as FIMP) and Case~II ($S$ as WIMP, $\chi$ as FIMP)---with substantially different coupling ranges, both of which are shown in the table.
\section{Summary and Outlook}\label{sec:sum}

In this work, we have investigated the FIMP-FIMP regime of a two-component dark matter model with a $Z_2\times Z_4$ symmetry. The model contains a singlet scalar $S$ and a Majorana fermion $\chi$ as dark matter candidates, where the fermion mass is generated through the symmetry-breaking relation $m_\chi=y_{sf}v_0$. We have shown that the ultra-feeble portal coupling required for the production of the scalar dark matter component is correlated with the freeze-in dynamics of the fermionic component rather than being an arbitrary choice.

In particular, reproducing the correct relic abundance requires a tiny Yukawa coupling $y_{sf}$ for $\chi$, which naturally leads to a large symmetry-breaking scale $v_0$. The resulting scale hierarchy induces an ultra-feeble portal coupling $\lambda_{ds}$ controlling the production of $S$. Through a systematic analysis of the FIMP-FIMP parameter space, we find that $\lambda_{ds}$ can reach values in the range $10^{-25}\lesssim\lambda_{ds}\lesssim10^{-13}$.
This demonstrates that extremely small dark matter interactions can arise from the internal structure of the dark sector rather than being imposed by hand.
We have also examined the contribution from gravitational freeze-in and identified it as an irreducible production mechanism at extremely small portal couplings. Therefore, the FIMP-FIMP regime possesses a natural lower boundary beyond which gravitational effects become increasingly relevant.

Our results provide a new perspective on ultra-feeble dark matter interactions in multi-component dark matter frameworks. Instead of treating tiny couplings as arbitrary inputs, the observed relic abundance, symmetry-breaking dynamics, and freeze-in production mechanism can together generate a hierarchical structure of interactions. Future studies may explore the implications of such ultra-feeble dark sectors for early-Universe cosmology, structure formation, and possible connections with ultraviolet completions of dark matter models.

\begin{acknowledgments}
\noindent
 Hao Sun is supported by the National Natural Science Foundation of China (Grant No.12075043, No.12147205). XinXin Qi is supported by the National Natural Science Foundation of China (Grant
No.12447162).
\end{acknowledgments}

\bibliography {v1}
\end{document}